\documentclass[%
 reprint,
superscriptaddress,
 amsmath,amssymb,
 pra,
]{revtex4-2}

\usepackage{graphicx}
\usepackage{dcolumn}
\usepackage{bm}
\usepackage[utf8]{inputenc}
\usepackage{floatrow}
\usepackage{threeparttable}

\newcommand{\bb}{\mathbf}
\newcommand{\dd}{\,\mathrm{d}}

\begin{document}

\title{The Microscopic Ampère formulation for the electromagnetic force density in linear dielectrics}

\author{B. Anghinoni}
\email{brunoanghinoni@gmail.com}
\affiliation{Department of Physics, Universidade Estadual de Maring\'a, Maring\'a, PR 87020-900, Brazil}
\author{M. Partanen}
\email{mikko.p.partanen@aalto.fi}
\affiliation{Department of Electronics and Nanoengineering, Aalto University, 00076 Aalto, Finland}
\author{N. G. C. Astrath}
\email{ngcastrath@uem.br}
\affiliation{Department of Physics, Universidade Estadual de Maring\'a, Maring\'a, PR 87020-900, Brazil}

\begin{abstract}
We present a detailed derivation of the electromagnetic force density and pressure in linear dielectric media according to the so-called Microscopic Ampère formulation, which considers the classical dipolar sources in matter along with the hidden momentum contribution. It is seen that, among the other formulations existing in the literature, our proposal is the only one simultaneously compatible with the experimental works reported to date and with the absence of magnetic monopoles in nature. A new radiation pressure equation for non-magnetic dielectrics under oblique illumination from p-polarized beams is also derived.
\end{abstract}

\maketitle

\section{Introduction}\label{sec:intro}
The complete knowledge of the electromagnetic forces acting inside matter when external fields are present stands as an unsolved problem in Physics. It is directly related to the centenary Abraham-Minkowski dilemma, which originally addressed the momentum of light inside linear dielectric media~\cite{Abraham,Minkowski,Brevik1979,Pfeifer2007,Anghinoni2022,Kemp2011,Milonni2010}. Although being a problem of fundamental Physics, this controversy has drawn much attention over the last few decades due to its close relation to optical manipulation techniques~\cite{Mansuripur2013b,Molloy2010,Gao2017,Neto2000,Ashkin1997,Shi2022,Li2020,Nieminen2007,Phillips1998}. Additionally, controlling  optomechanical effects are of great interest for the development of photonic devices~\cite{Yang2022,Chin2020,Wiederhecker2009,Partanen2022b} and for optofluidic technology~\cite{Psaltis2006,Monat2007,Garnier2003}, for example. 

The force density on charged matter is unambiguously given by the Lorentz force law,
\begin{equation}
    \bb f = \rho \bb E + \bb J \times \bb B,
\end{equation}
where $\rho$ is the electric charge density, $\bb J$ is the electric current density, $\bb E$ is the electric field and $\bb B$ is the magnetic induction field. The different force densities in each formulation existing in the literature can then be attributed to different modeling of the electromagnetic sources $\rho$ and $\bb J$. These sources are, of course, not arbitrary -- they must be consistent with Maxwell's equations and with the conservation of electric charge.

Historically, besides the two formulations that name the Abraham-Minkowski dilemma, the specialized literature discusses three additional main formulations for the electromagnetic force density generated in dielectrics due to the application of external fields, namely, the formulations from Einstein and Laub~{\cite{EL}}, Ampère and Chu~\cite{Chu}. It can be seen that all of them present their own problems~\cite{Anghinoni2022}. First, Minkowski's formulation totally neglects the bound charges inside the material~\cite{Milonni2010}. In fact, had he used the correct charge density, he would have obtained Ampère's conventional formulation without the hidden momentum contribution (see Sec.~\ref{sec:HM}). As Abraham's formulation only differs from Minkowski's in the momentum density, it should share this same problem. 
Ampère's formulation does adopt the appropriate classical microscopic model, i.e., the electric sources are given as ideal dipoles and the magnetic sources as tiny current loops~\cite{Jackson1977}. This formulation works properly if we are interested only in the movement of the center of mass of rigid bodies. However, as only the macroscopic effects of polarizations and magnetizations are considered, Ampère's formulation is not expected to correctly describe the microscopic force distribution inside materials~\cite{Griffiths2015}. 
The Einstein-Laub formulation, on its turn, has been built considering the usual dipolar approximation for the electromagnetic fields inside matter, but adopts an incorrect model for the microscopic magnetization mechanism, as it assumes the existence of magnetic monopoles, which have never been observed and showed disagreement with experiments~\cite{Jackson1977,Hughes1951,Mezei1986}. For non-magnetic materials, its related force density should be correct -- however, there would still remain strong theoretical issues, like the absence of Lorentz invariance of the electromagnetic stress-energy tensor~\cite{Kemp2017}. 
Although attributing different contributions to light and matter, Chu's formulation follows the same microscopic model of Einstein and Laub.  At last, every formulation except Einstein-Laub's and Chu's do not account naturally for electro- and magnetostriction effects, which are quadratic effects on the fields tending to compress the medium towards regions of higher field intensity~\cite{LL}. A table summarizing the different force densities from these formulations can be found in Appendix~\ref{sec:app_tab}. For a more detailed discussion on the different electromagnetic formulations, see for example Refs.~\cite{Anghinoni2022,Kemp2011,Brevik1979,Penfield1967}.


In the context just presented, it seems natural to consider a formulation that arises from the classical charge and current distributions related to microscopic electric and magnetic dipoles. This has been attempted earlier in Refs.~\cite{Shevchenko2010,Shevchenko2011}, but in the absence of the hidden momentum contribution (discussed in detail in next section). An axiomatic approach was also reported~\cite{Mansuripur2009b}, where it is argued that no specific model for the electromagnetic sources in matter has to be adopted for a consistent description of electromagnetism. In this work, we employ the classical dipolar approximation for electromagnetic sources in matter to derive an electromagnetic force density that will be shown to present all the characteristics necessary to explain the existing experiments. In doing so, the enigma of the hidden momentum is also clarified and properly added to the force density.

This work is organized as follows. In Sec.~\ref{sec:HM} we discuss the essential concept of hidden momentum in the context of the Abraham-Minkowski problem. In Sec.~\ref{sec:charge} we describe the electromagnetic sources associated with ideal dipoles for both static and time-dependent cases. In Sec.~\ref{sec:f+p} we derive the electromagnetic force density related to these sources, as well as the radiation pressure. In Sec.~\ref{sec:exp} we compare the obtained results with many experimental investigations reported so far, followed by the discussions in Sec.~\ref{sec:disc}.
Lastly, in Sec.~\ref{sec:conc} we summarize our results.

\section{Hidden momentum}\label{sec:HM}

In 1967, Shockley and James~\cite{Shockley1967} identified a previously unrecognized source of linear momentum that should arise when a magnetic dipole moment $\bb m$ interacts with an electric field $\bb E$ -- even if both did not vary in time. This source became known as hidden momentum, and is given by
\begin{equation}\label{eq:HM}
    \bb p_{\mathrm{h}} = \frac{1}{c^2}(\bb m \times \bb E),
\end{equation}
where $c$ is the speed of light in vacuum. Since then, many works tried to properly interpret this puzzling term, which is inevitably tied to the Abraham-Minkowski controversy. Some authors claimed that it occurs as a relativistic effect in systems that are macroscopically at rest, but contains internally moving parts, such as a common electric circuit~\cite{Griffiths,Correa2020,Babson2009,KT2018} (see also Ref.~\cite{McDonald2020} and references therein). 
It was also shown that hidden momentum is necessary to keep the correct relativistic properties of energy, momentum and rest mass of a charge and current carrying body~\cite{Hnizdo1997}. Indeed, it was recently suggested that hidden momentum is a general relativistic concept, not exclusive to electromagnetic systems~\cite{HM}.

Apart from these historical conceptual issues, it was formally shown~\cite{Horsley2006} that starting from the conventional Quantum Electrodynamics (QED) Lagrangian for a point, spinless charged particle in relativistic motion and properly applying the center of mass-energy theorem there must be an extra momentum given by Eq.~(\ref{eq:HM}). Classically, the correct interpretation of hidden momentum is actually quite simple: a moving electric dipole develops a magnetic dipole. More specifically, this occurs when the electric dipole moment $\bb p$ of a particle moving with velocity $\bb v$, both measured in the laboratory frame, is Lorentz-transformed to the particle's rest frame~\cite{Sonnleitner2017}. To first order in $|\bb v|/c$, the new electric dipole moment is~\cite{Hnizdo2012} $\bb p= \bb p' + \bb v \times \bb m'/c^2$, where $\bb p'$ and $\bb m'$ are the particle's electric and magnetic dipole moments, respectively, in its rest frame. The hidden momentum contribution then comes exactly from this last term.

In 1984 Aharonov and Casher showed in a seminal work \cite{Aharonov1984} that the hidden momentum arises as a topological quantum effect when describing the interaction between a charged particle and a magnetic moment, where they obtained Eq.~(\ref{eq:HM}) as a non-relativistic limit of the Dirac equation -- in this context, the hidden momentum is also known in the literature as Aharonov-Casher interaction. In analogy with the Aharonov-Bohm effect \cite{Aharonov1959}, this interaction does not necessarily generate a force, but introduces a phase shift in the wave function of the system, which has already been observed -- see Refs.~\cite{Sangster1995,Pop2012} for example.

Notice that the symmetry inherent to Maxwell's equations requires the existence of an effect analogous to hidden momentum for magnetic dipoles, i.e., an effect due to moving magnetic dipoles generating electric dipoles. This indeed takes place and is known in the literature as Röntgen interaction. Its momentum is given by $\bb p_{\mathrm{R}} = -\bb p \times \bb B$ 
and can also be rigorously obtained from the QED framework~\cite{Horsley2006,Baxter1993,Lembessis1993}. This interaction can also generate a topological phase~\cite{Wilkens1994}, but, to our knowledge, such effect has not been observed yet.

Although hidden momentum has certainly been subject of more intense discussions in the literature, both interactions presented here are of equal importance, and they are expected to take essential part in the eventual resolution of the Abraham-Minkowski problem. They are known to arise when the center of mass-energy of the system is regarded as a dynamic variable. However, the Röntgen term appears naturally even in non-relativistic derivations (see Refs.~\cite{Lembessis1993,Stenholm1986} for example), while the hidden momentum necessarily requires a relativistic treatment, as shown in Ref.~\cite{Horsley2006}. The Röntgen interaction and hidden momentum contribute to the electromagnetic force density as, respectively, $\bb f_{\mathrm{R}}=\dd (\bb P \times \bb B)/ \dd t$ and $\bb f_{\mathrm{h}}=-\dd(\bb M \times \bb E /c^2)/ \dd t$, where $\bb P$ and $\bb M$ are the medium's polarization and magnetization, respectively, and the minus signs added to both equations stem from the fact that the force densities are generated due to the fields losing their momentum. The latter contribution to the force density is important even in systems with non-relativistic velocities~\cite{Hnizdo1997}, and so it must be properly added \textit{ad hoc} in the results from non-relativistic derivations.

\section{Dipolar sources}\label{sec:charge}

In this section we present the microscopic electromagnetic sources associated to a classical point electric and magnetic dipole. The dipole is composed of two point charges of opposite value $\pm q$ separated by a distance $|\bb d|$. The chosen inertial reference frame is the rest frame of the dipole -- consequently, the velocity of the dipole's center of mass is taken as zero. The dipole moments will first be considered as static and later to be time-dependent.

\subsection{Static dipole}

The charge density associated to the classical dipole just described at a point $\bb r$ is given by~\cite{Zangwill2013,Griffiths2015,Jackson1999}
\begin{equation}\label{eq:rho}
    \rho (\bb r) =-(\bb p \cdot \bm \nabla)\delta^3(\bb r),
\end{equation}
where $\bb p=q\bb d$ is the electric dipole moment and $\delta^3(\bb r)$ is the three-dimensional Dirac delta function, i.e., $\delta^3(\bb r) = \delta(x)\delta(y)\delta(z)$. On its turn, the current density in this model is
\begin{equation}\label{eq:J}
    \bb J (\bb r) = -\bb m \times \bm\nabla\delta^3(\bb r),
\end{equation}
where $\bb m = (1/2)\int \bb r \times \bb J \dd^3 \bb r$ is the magnetic dipole moment. 
Notice that a formal calculation would require the consideration of dynamic dipoles (i.e., both $\bb p$ and $\bb m$ generally time-dependent) from the start because their acceleration can generate radiation-related terms. 
Besides, the concept of retarded time must also be included to assure physical causality. These requirements are addressed in the next section.

\subsection{Time-dependent dipole}

It is well-known that in the Lorenz gauge the electromagnetic potentials $\varphi$ and $\bb A$ are described by non-homogeneous wave equations whose formal solutions are~\cite{Jackson1999} 
\begin{equation}\label{eq:phi}
    \varphi(\bb r,t) = \frac{1}{4\pi\varepsilon_0}\int \frac{\rho(\bb r',t_\mathrm r)}{|\bb r-\bb r'|} \dd^3\bb r'
\end{equation}
and
\begin{equation}\label{eq:A}
    \bb A (\bb r, t) = \frac{\mu_0}{4\pi}\int \frac{\bb J(\bb r',t_\mathrm r)}{|\bb r-\bb r'|}\dd^3\bb r',
\end{equation}
where $t_\mathrm r = t - |\bb r-\bb r'|/c$ is the retarded time. Under the dipolar approximation, the electromagnetic potentials of dynamic point dipoles are given by \cite{Griffiths2011}
\begin{equation}\label{eq:phi_dd}
    \varphi(\bb r, t) = \frac{1}{4\pi\varepsilon_0}\left[\bb p(t_0) + \frac{r}{c}\dot{\bb p}(t_0) \right]\cdot\frac{\hat{\bb r}}{r^2}
\end{equation}
and
\begin{equation}\label{eq:A_dd}
    \bb A(\bb r,t) = \frac{\mu_0}{4\pi}\left[\frac{\dot{\bb p}(t_0)}{r}+\frac{\bb m(t_0)\times\hat{\bb r}}{r^2}+\frac{\dot{\bb m}(t_0)\times\hat{\bb r}}{c r}\right],
\end{equation}
where $\bb p(t)$ and $\bb m(t)$ are the now time-dependent electric and magnetic dipole moments, respectively, and $t_0 = t-r/c$ is the retarded time at the origin.

The electromagnetic fields are given in terms of the potentials as $\bb E = -\bm \nabla \varphi - \partial_t \bb A$ and $\bb B = \bm \nabla \times \bb A$. We can use Gauss' Law and Ampère-Maxwell's law to obtain the time-dependent charge and current densities as $\rho(\bb r, t) = -\varepsilon_0 \nabla^2\varphi -\varepsilon_0 \bm \nabla \cdot \partial_t \bb A$ and
$\bb J(\bb r,t) = \mu_0^{-1} \bm \nabla \times \bm \nabla \times \bb A$.
This is, however, not a very convenient procedure because the implicit dependence of $t_0$ in $\bb r$ makes the calculation very difficult. 
Alternatively, it is sufficient to show that the sources adopted in the last section, 
when promoted to time-dependent, generate the correct electromagnetic potentials given in Eqs.~(\ref{eq:phi_dd}) and~(\ref{eq:A_dd}) when calculated using Eqs.~(\ref{eq:phi}) and~(\ref{eq:A}). We start by calculating the electric potential due to $\rho(\bb r,t) = - (\bb p(t)\cdot\bm\nabla)\delta^3(\bb r)$ as
\begin{equation}
    \varphi (\bb r,t) = -\frac{1}{4\pi\varepsilon_0}\int \frac{(\bb p(t_\mathrm r)\cdot\bm\nabla')\delta^3(\bb r')}{|\bb r-\bb r'|} \dd^3\bb r'.
\end{equation}
Using the property $\int f(x)\delta'(x-x_0)\dd x = -f'(x_0)$, we have
\begin{equation}\label{eq:aa}
     \varphi (\bb r,t) = \frac{1}{4\pi\varepsilon_0}\bm\nabla'\cdot\left(\frac{\bb p(t_\mathrm r)}{|\bb r-\bb r'|}\right)_{\bb r'=0}.
\end{equation}
Due to the implicit dependence of $t_\mathrm r$ on $r$, we have $\bm \nabla' \cdot \bb p(t_\mathrm r)\vert_{\bb r'=0} = \tilde{\bm \nabla}'\cdot \bb p(t_0) +(\hat{\bb r}/c)\cdot\dot{\bb p}(t_0)$, where $\tilde{\bm \nabla}'$ denotes the nabla operator acting only on the spatial coordinates. Specifically, in the point dipole approximation we have $\tilde{\bm \nabla}'\cdot \bb p(t_0) =0$ as $\bb p$ does not depend on $\bb r$, so that the electric potential in Eq.~(\ref{eq:aa}) is reduced to Eq.~(\ref{eq:phi_dd}), as expected.

To calculate the vector magnetic potential $\bb A$, we first notice that when $\bb p$ is time-dependent there is an extra term $\dot{\bb p}(t)\delta^3(\bb r)$ in $\bb J$ originating from the continuity equation, $\partial_t \rho = -\bm\nabla\cdot\bb J$, where the over-dot denotes time derivative. Therefore, $\bb J(\bb r, t)=\dot{\bb p}(t)\delta^3(\bb r)-(\bb m(t) \times \bm \nabla)\delta^3(\bb r)$, and the vector magnetic potential is
\begin{equation}
    \bb A(\bb r,t) = \frac{\mu_0}{4 \pi}\int\frac{\dot{\bb p}(t_\mathrm r)\delta^3(\bb r')-(\bb m(t_\mathrm r)\times\bm\nabla')\delta^3(\bb r')}{|\bb r-\bb r'|}\dd^3\bb r'.
\end{equation}
The first term is trivially integrated to $(\mu_0/4\pi r)\dot{\bb p}(t_0)$. The second term is analogous to Eq.~(\ref{eq:aa}), with $\bb p \to \bb m$ and divergence operator $\to$ curl operator, resulting in Eq.~(\ref{eq:A_dd}) and completing our calculation. Thus, we have shown that if the dipolar approximation can be suitably applied, the dynamical microscopic electromagnetic sources can be described as
\begin{equation}\label{eq:charge}
    \rho(\bb r, t) = -(\bb p(t)\cdot\bm\nabla)\delta^3(\bb r)
\end{equation}
and
\begin{equation}\label{eq:current}
    \bb J(\bb r, t) =\dot{ \bb p}(t)\delta^3(\bb r)-(\bb m(t) \times \bm \nabla )\delta^3(\bb r).
\end{equation}
Consequently, by employing the Lorentz force density to these sources we should be able to obtain the appropriate force density distribution inside matter within the adopted approximations. 

Lastly, we notice that the retarded time $t_\mathrm r$ is known to be related to general scattering phenomena inside matter according to the Ewald-Oseen extinction theorem~\cite{Mansuripur98,BornWolf}. As we adopt here the common effective, continuum description of dielectrics through the use of the macroscopic parameters $\varepsilon$ and $\mu$, the effects of the retarded time are already implicitly contained in the resultant fields. Thus, in our model the fields and forces can be evaluated at the regular time $t$.


\section{Electromagnetic force density and pressure}\label{sec:f+p}

We consider electromagnetic fields inside dielectric materials within the optical bandwidth. To calculate the electromagnetic force density and pressure, we suppose that at this optical length scale the microscopic sources are well described by the dipolar approximation, as discussed in the last section. 

\subsection{Force density}
The force acting on charged matter is unambiguously given by the continuous version of the Lorentz force law,
\begin{equation}\label{eq:Lorentz}
    \bb F = \int_{\delta V}\left(\rho \bb E + \bb J \times \bb B\right) \dd^3 \bb r,
\end{equation}
which after some algebra (see calculations in Appendix~\ref{sec:app_f}) leads to the force density in the Microscopic Ampère (MA) formulation,
\begin{eqnarray}\label{eq:f2_}
    \bb f_{\mathrm{MA}} = \frac{\varepsilon_0(\varepsilon_{\mathrm{r}}-1)}{2}\bm\nabla |\bb E|^2
     +(\mu_{\mathrm{r}}-\!1)|\bb H|^2\bm\nabla\mu
     \nonumber\\
     +\!\frac{\mu(\mu_{\mathrm{r}}-1)}{2}\bm\nabla|\bb H|^2 
 +\frac{\partial}{\partial t}\left(\bb P \!\times \!\bb B\right). 
\end{eqnarray}
Here, by ``force density'' we mean the electromagnetic force acting on a small volume of the dielectric, $\delta V$, which is microscopically large (i.e., encompasses a large number of dipoles), but is still much smaller than the dielectric's macroscopic volume. Also,
$\varepsilon_{\mathrm{r}}=\varepsilon/\varepsilon_0$ is the relative permittivity and $\mu_{\mathrm{r}}=\mu/\mu_0$ the relative permeability, while $\bb H = \mu^{-1} \bb B$ is the magnetic field. The last term in Eq.~(\ref{eq:f2_}) is the Röntgen interaction, which naturally appeared in our non-relativistic derivation for a dipole at rest, as anticipated in Sec.~\ref{sec:HM}.
If we employed a relativistic derivation for a moving dipole from the very beginning, there would be an extra contribution -- the hidden momentum -- as shown in Ref.~\cite{Horsley2006}. An alternative
non-relativistic derivation for the force density (in non-magnetic media) where the dipole is moving can be found in Ref.~\cite{Stenholm1986}.

As discussed in Sec.~\ref{sec:HM}, when switching to the laboratory frame it is necessary to add the hidden momentum contribution to Eq.~(\ref{eq:f2_}) as $\bb f_{\mathrm{h}}\approx \partial_t(\bb E \times \bb M)/c^2$. The time derivative approximation $\dd /\dd t \approx \partial/\partial t$ takes place because the dipole's velocity (as measured in the laboratory frame) is much smaller than $c$. The force density is then
\begin{eqnarray}\label{eq:f22}
    \bb f_{\mathrm{MA}} = \frac{\varepsilon_0(\varepsilon_{\mathrm{r}}-1)}{2}\bm\nabla |\bb E|^2
     +(\mu_{\mathrm{r}}-\!1)|\bb H|^2\bm\nabla\mu
      \nonumber\\
 +\!\frac{\mu(\mu_{\mathrm{r}}-1)}{2}\bm\nabla|\bb H|^2+\frac{n^2-1}{c^2}\frac{\partial}{\partial t}\left(\bb E \!\times \!\bb H\right),  
\end{eqnarray}
where $n=\sqrt{\varepsilon_{\mathrm{r}}\mu_{\mathrm{r}}}$ is the refractive index and it is assumed that $\varepsilon$ and $\mu$ do not depend on time. A simple rearrangement of the gradients as products yields, at last, the main result of our work,
\begin{eqnarray}\label{eq:f3}
    \bb f_{\mathrm{MA}} = \frac{1}{2}\bm\nabla\left(\bb P \!\cdot\! \bb E \right)
 \!+\!\frac{1}{2}\bm\nabla\left(\bb M \!\cdot\! \bb B \right)
  \!-\!\frac{1}{2}|\bb E|^2\bm\nabla\varepsilon   \nonumber \\
  \!-\!\frac{1}{2}|\bb H|^2\bm\nabla\mu
  \!+\!\frac{n^2\!-\!1}{c^2}\frac{\partial}{\partial t}\left(\bb E \!\times\! \bb H\right). 
\end{eqnarray}
This force density equation is given in the laboratory frame. It is valid for linear, isotropic inhomogeneous media, with $\varepsilon$ and $\mu$ independent of time, i.e., no dispersion. The presence of free sources would generate the extra terms $\rho_{\mathrm{f}} \bb E$ and $\bb J_{\mathrm{f}} \times \bb B$ in Eq.~(\ref{eq:Lorentz}), and can be included if necessary. The first and second terms can be assigned as the electrostriction and magnetostriction force densities, respectively. 
The third and fourth terms are the usual Abraham-Minkowski force, which occur in non-homogeneous regions, and the last term is the well-known Abraham force. This equation contemplates almost every aspect of the reported experiments (as will be discussed in Sec.~\ref{sec:exp}), and arises naturally from a clear and simple microscopic model, with no need for phenomenological approaches. An alternative derivation of Eq.~(\ref{eq:f3}) for homogeneous media using the Lagrangian approach is found in Appendix~\ref{sec:Lag}.


\subsection{Radiation pressure}\label{sec:surf_p}

The electromagnetic pressure, widely known in the literature as radiation pressure, is here denoted by $\mathcal{P}_{\mathrm{rad}}$ and calculated at a flat dielectric interface assuming a laser beam with azimuthal symmetry about the propagation axis. The deformations induced by radiation pressure in dielectric liquids are typically bulges of height of order 10 nm~\cite{Astrath2014,Astrath2022,Chaudhary2019,Verma2017,Verma2015}, rapidly decreasing over a length of about one beam waist, which is usually of order 100 $\mu$m. Thus, considering the interface flat even when the fields are acting on it is certainly a good approximation. 

\begin{figure}
    \centering
    \includegraphics[width=0.9\textwidth]{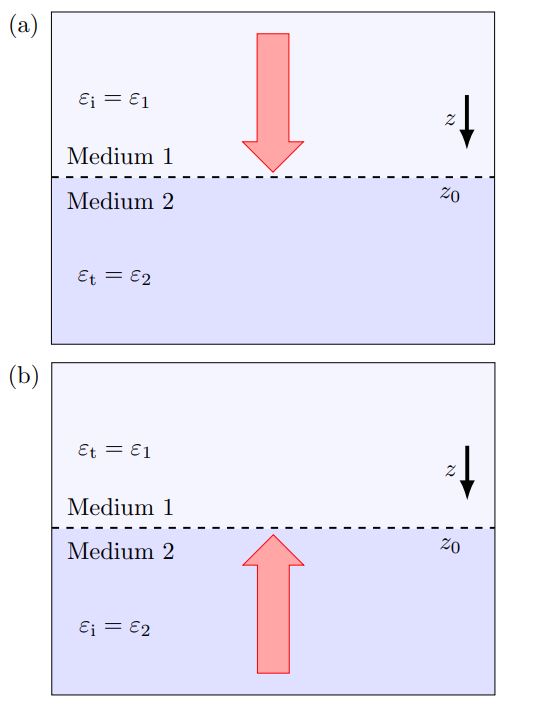}
  \caption{Convention for permittivities used in radiation pressure equation, Eq.~(\ref{eq:P_rad}), according to beam incidence direction. In (a), the beam propagates in $z$ direction, while in (b) the propagation direction is reversed.}\label{fig:interface2}
\end{figure}

As an isotropic quantity, the radiation pressure has two contributions in this case: one due to the discontinuity of refractive index, $\bb P$ and $\bb M$ in the direction normal to the interface and another one due to the difference in the radial forces (electro- and magnetostriction effects) in each medium. The first contribution can be obtained from Eq.~(\ref{eq:f3}) by properly integrating the normal component of the force density. For example, for a monochromatic beam propagating in $z$ direction incident from a non-magnetic, linear and isotropic dielectric medium into a flat interface at $z_0=0$, as illustrated in Fig.~\ref{fig:interface2}~(a), we have
\begin{eqnarray}\label{eq:Pz}
    \mathcal{P}_z &=& \lim_{\delta \to 0}\int_{-\delta}^{+\delta}f_z \dd z \nonumber \\
      &=& \lim_{\delta \to 0}\int_{-\delta}^{+\delta}\left[\frac{1}{2}\frac{\partial}{\partial z}\left(\bb P \cdot \bb E\right)-\frac{1}{2}|\bb E|^2\frac{\partial \varepsilon}{\partial z}\right] \dd z \nonumber \\
    &=&\left[\frac{1}{2}(\bb P\cdot\bb E)\right]^{z=0^{+}}_{z=0^{-}}-\frac{\varepsilon_2-\varepsilon_1}{2}\left( |\bb E|^2_{\mathrm{avg}}\right).
\end{eqnarray}
We assumed in Eq.~(\ref{eq:Pz}) there are no free charges at the dielectric interface. Here, $|\bb E|^2_{\mathrm{avg}}$ is the average of the squared electrical field magnitude across the interface and the last term of Eq.~(\ref{eq:f3}) has been averaged out. 

The second term of the radiation pressure at the interface is related to the radial direction, and for normal incidence is given by 
\begin{eqnarray}\label{eq:Pr}
    \mathcal{P}_r &=& -\left[\int f_r \dd r\right]_{z=0^{-}}^{z=0^{+}}
      = -\left[\frac{1}{2}\left(\bb P \cdot \bb E\right)\right]_{z=0^{-}}^{z=0^{+}} 
\end{eqnarray}
where the same assumptions were used again and the function inside the brackets was implicitly calculated at fixed $r$.

Summing the two contributions, we have the radiation pressure at the flat dielectric interface as
\begin{equation}\label{eq:P_AM}
    \mathcal{P}_{\mathrm{rad}}=\mathcal{P}_z+\mathcal{P}_r=-\frac{\left(\varepsilon_2-\varepsilon_1\right)}{2} |\bb E|^2_{\mathrm{avg}}.
\end{equation}
Applying Maxwell's equations boundary conditions, we have then
\begin{equation}\label{eq:P_rad}
    \mathcal{P}_{\mathrm{rad}}=-\frac{\left(\varepsilon_2-\varepsilon_1\right)}{2}\left[E_x^2+E_y^2+\left(1+\frac{\varepsilon_{\mathrm{t}}^2}{\varepsilon_{\mathrm{i}}^2} \right)\frac{E_{z,\mathrm{t}}^2}{2}\right].
\end{equation}
Here $E_{z,\mathrm{t}}$ is the transmitted field component normal to the interface. As this component is not continuous across the interface, we averaged its squared magnitude with a simple arithmetic mean. The tangential components $E_x$ and $E_y$ are continuous across the interface, and therefore do not need the subscript indicating the current medium. Notice a very important subtlety introduced by this equation: the permittivities $\varepsilon_1$ and $\varepsilon_2$ are related to the direction of $z$, which in our convention always points from medium 1 to medium 2. The gradient $\bm \nabla \varepsilon$ is calculated accordingly, resulting in the term outside the brackets in Eq.~(\ref{eq:P_rad}). On the other hand, the permittivities $\varepsilon_{\mathrm{i}}$ and $\varepsilon_{\mathrm{t}}$ are related to the beam propagation direction, i.e., they refer to incident and transmitted components, respectively. Thus, if the beam is propagating in the $z$ direction, we have $\varepsilon_{\mathrm{i}}=\varepsilon_1$ and $\varepsilon_{\mathrm{t}}=\varepsilon_2$, as in Fig.~\ref{fig:interface2}~(a); if the propagation direction is reversed, we have $\varepsilon_{\mathrm{t}}=\varepsilon_1$ and $\varepsilon_{\mathrm{i}}=\varepsilon_2$, as in Fig.~\ref{fig:interface2}~(b).

By using the conventions just described, Eq.~(\ref{eq:P_rad}) is then valid for any beam polarization, incidence angle and propagation direction. Specifically, for normal incidences we have $E_z^2 \ll E_x^2+E_y^2$, even for typical focused Gaussian beams. This leads to $\mathcal{P}_{\mathrm{rad}} \approx -(\varepsilon_2-\varepsilon_1) (E_x^2+E_y^2)/2$, which is the widely known Abraham-Minkowski radiation pressure. In fact, as the electrostriction effects are seen to cancel at the interface, Eq.~(\ref{eq:P_AM}) could also be obtained through Abraham's force density, $\bb f_{\mathrm{Ab}}=-|\bb E|^2\bm\nabla\varepsilon/2$ -- however, electrostriction plays an important role in the stability of fluids, as will be discussed in Sec.~\ref{sec:exp_p}.

We see that keeping the incidence as normal and reversing beam propagation direction would generate the same radiation pressure equation. For oblique beam incidences, we must properly account the different reflection and transmission coefficients for each polarization.
We will consider the beam is locally a plane wave -- which is a good approximation, since even for focused Gaussian beams the field components in directions other than the polarization one are typically negligible. In this condition, we can apply Fresnel equations to describe the reflected and transmitted field amplitudes. First, for s polarization, the field is by definition perpendicular to the plane of incidence. In this case, we have in our convention $E_z = 0$, which generates
\begin{equation}\label{eq:P_s}
    \mathcal{P}_{\mathrm{rad}}^{(\mathrm{s})} = -\frac{\left(\varepsilon_2-\varepsilon_1\right)}{2}t_{\mathrm{s}}^2 (\theta_{\mathrm{i}})E_0^2,
\end{equation}
where $E_0$ is the field amplitude, $t_{\mathrm{s}}$ is the transmission coefficient for s polarization and $\theta_{\mathrm{i}}$ is the incident angle relative to the interface's normal direction.

For p polarized beams, we have a non-zero normal component, so that the radiation pressure becomes
\begin{eqnarray}\label{eq:P_p}
 \mathcal{P}_{\mathrm{rad}}^{(\mathrm{p})} = -\frac{\left(\varepsilon_2-\varepsilon_1\right)}{2}E_0^2\left[t_{\mathrm{p}}^2 (\theta_{\mathrm{i}})\cos^2\theta_{\mathrm{t}} \right. \nonumber \\ \left.+\frac{(1+r_{\mathrm{p}}(\theta_{\mathrm{i}}))^ 2\sin^2\theta_{\mathrm{i}}+t_{\mathrm{p}}^2(\theta_{\mathrm{i}})\sin^2\theta_{\mathrm{t}}}{2}\right],
\end{eqnarray}
where $t_{\mathrm{p}}$ and $r_{\mathrm{p}}$ are the transmission and reflection coefficients for p polarization, respectively, and $\theta_{\mathrm{t}} = \sin^{-1}((n_1/n_2)\sin\theta_{\mathrm{i}})$ is the transmitted (refracted) angle. The terms divided by two within the brackets correspond exactly to the spatial average of $E_z^2$ across the interface.

\section{Comparison to experiments}\label{sec:exp}

As we have seen in Sec.~\ref{sec:f+p}, the MA formulation naturally accounts for the electro- and magnetostriction effects, presents a radiation pressure of the Abraham-Minkowski form and an Abraham-type momentum density. To the best of our knowledge, there is no force density in the literature valid for optical regime that simultaneously presents all these characteristics without including phenomenological approaches. These fundamental properties will be used in this section to analyze the main existing experiments related to electromagnetic force density. 
We emphasize that the great majority of these experiments were quantitatively described in terms of one of the previously known electromagnetic formulations, either in the original or in subsequent works. Our objective here is to show if and how MA formulation can also be applied to interpret them.
For better organization, these experiments are grouped in four categories: radiation pressure experiments, photon momentum experiments, bulk force experiments and total force experiments. 

\subsection{Radiation pressure experiments}\label{sec:exp_p}

The surface deformation of water under normal laser incidence was successfully explained using the radiation pressure given in Eq.~(\ref{eq:P_rad}), both in old and recent measurements \cite{Ashkin1973,Astrath2014,Capeloto2015} -- specifically, in Ref.~\cite{Ashkin1973} the reversed beam propagation direction was also considered. An interface of different fluids close to the critical point was studied in Ref.~\cite{Casner2001} and the observed surface deformations were also well described by Eq.~(\ref{eq:P_rad}).

The radiation pressure for oblique incidence adopted in the literature is~\cite{Casner2003,Casner2003b,Borzdov1993,Hallanger2005,Girot2019}
\begin{equation}\label{eq:P_lit}
    \mathcal{P}_{\mathrm{rad}} = \frac{n_\mathrm{i} I}{c} \cos^2\theta_{\mathrm{i}}\left[1+R(\theta_\mathrm{i})-\frac{\tan\theta_{\mathrm{i}}}{\tan\theta_{\mathrm{t}}}T(\theta_{\mathrm{i}})\right],
\end{equation}
where $n_\mathrm{i}$ is the incidence medium's refractive index, $I$ is the beam intensity and $R$ and $T$ are the interface's reflectance and transmittance, respectively. This equation contemplates both polarizations in a single equation by using the appropriate $R$ and $T$, and has been applied to explain the experiments reported in Refs.~\cite{Schaberle2019,Verma2015,Verma2017,Casner2003,Casner2003b,Girot2019}.
However, this equation does not account properly for the discontinuity of the normal field at the interface for p polarization, as shown in Appendix~\ref{sec:app_P}. 

\begin{figure}
    \centering
    {\includegraphics[width=1.00\textwidth]{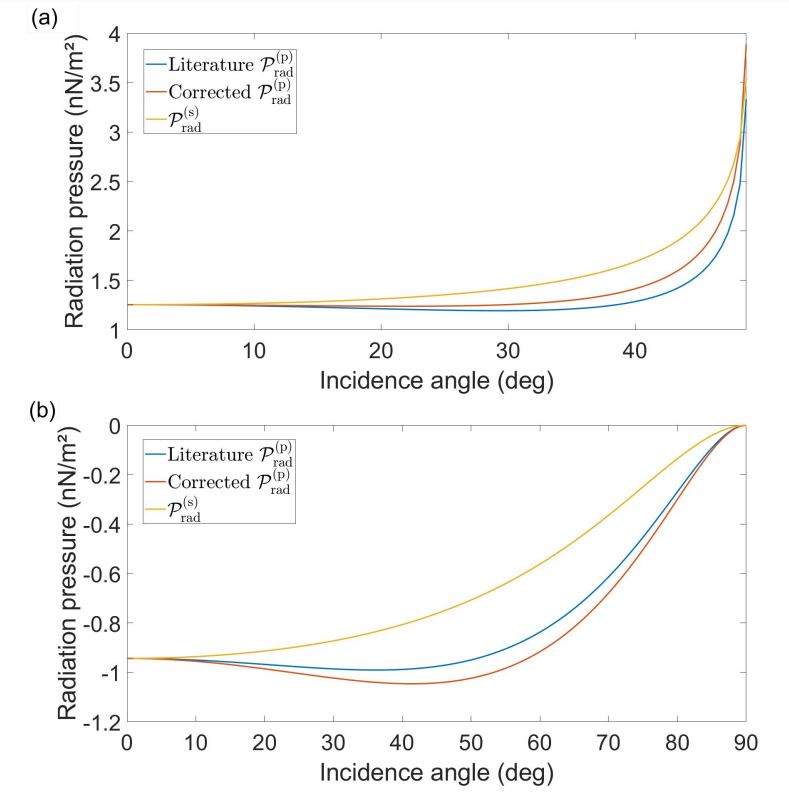}}
   \caption{Radiation pressure for oblique incidence and p-polarized beams according to the literature, Eq.~(\ref{eq:P_lit}), and to our suggestion, Eq.~(\ref{eq:P_p}). In (a) we have an water-to-air incidence and the radiation pressure is calculated for angles smaller than the critical angle for total internal reflection, $\theta_{\mathrm{c}}\approx 48.77\,^{\circ}$. In (b) we have an air-to-water incidence. The result for s polarization is also shown for completeness. Beam intensity is 1.0 W/m². The relative permittivities used for water and air were 1.769 and 1.0, respectively.}\label{fig:P_obl}
\end{figure}

For an air-water interface, Fig.~(\ref{fig:P_obl}) shows the radiation pressures from Eqs.~(\ref{eq:P_p}) and (\ref{eq:P_lit}) for both incidence directions. The behavior is qualitatively the same, but the magnitude of the corrected version is about 5--10\% larger. At first sight, our result seems compatible with the deformations observed in Refs.~\cite{Verma2015,Verma2017,Schaberle2019}, but further investigations are necessary. Specifically, we notice that in Refs.~\cite{Verma2017,Verma2015} the ellipsoidal character of the incident beam's cross section at the interface under different incidence angles was apparently not taken into account -- a fact that can significantly alter the employed theory.

Notice that when considering a fluid incompressible we are assuming any information about deformations in the fluid propagates instantaneously -- thus, no transient response is present. Indeed, at equilibrium, applying the divergence operator to the Navier-Stokes equation for an incompressible fluid at rest yields~\cite{Batchelor}
\begin{equation}\label{eq:Poisson}
    \nabla ^2 \mathcal{P} = \bm \nabla \cdot \bb f_{\mathrm{em}},
\end{equation}
where $\bb f_{\mathrm{em}}$ is the electromagnetic body force and $\mathcal{P}$ is the fluid's pressure. This is an elliptic partial differential equation for $\mathcal{P}$, known as Poisson's equation. It is well-known to possess unique solutions (up to an additive constant) for a very broad class of boundary conditions. Thus, in this situation, the Abraham-Minkowski pressure, Eq.~(\ref{eq:P_rad}) with $E_z = 0$, arises naturally as a boundary condition uniquely related to the divergence of the body force density from Eq.~(\ref{eq:f3}). This is in accordance with the fact that striction effects play a significant role in the stability of fluids~\cite{Ellingsen2011,Casner2003,Brueckner1966,Brevik1982}. On the other hand, if the fluid develops a position-dependent velocity field, the pressure at the surface can not be uniquely related to the body force anymore, as there will be another source term in Eq.~(\ref{eq:Poisson}). In fact, the pressure (and consequently the surface deformation) can even change signs, as shown in Ref.~\cite{Leonhardt2014}. This is a possible explanation to the Abraham-type deformation of a free fluid surface reported in Ref.~\cite{Ab1}.

\subsection{Photon momentum experiments}\label{sec:pme}

The Abraham force (last term of Eq.~(\ref{eq:f3})) is known to be related to a photon momentum proportional to $n^{-1}$~\cite{Milonni2010,Anghinoni2022,Barnett2010}, which is usually referred to as Abraham-type momentum in the context of the Abraham-Minkowski controversy, while the Minkowski-type momentum is proportional to $n$.
The recoil of a submerged mirror due to radiation pressure was measured twice~\cite{Jones1954,Jones1978} and the results were directly proportional to the refractive index of the background dielectric media -- i.e., of Minkowski's form. This can effectively be explained by the Doppler-shifted recoil of the mirror, while the field retains Abraham's form of momentum \cite{Milonni2010}: the incident photon has a momentum proportional to $n^{-1}$, but, due to Doppler's effect, the mirror's recoil is described by a momentum transfer linear in $n$ if the recoil velocity is non-relativistic. Therefore, by considering this interaction of field and matter, the results from Refs.~\cite{Jones1954,Jones1978} are in accordance with MA force density.
Alternatively, this can also be described by the mass-polariton quasi-particle model~\cite{MP1}, where a light-induced atomic mass density wave carries the difference of the Minkowski and Abraham momenta as discussed in Sec.~\ref{sec:disc}.
This explanation also applies to Ref.~\cite{Campbell2005}, where the recoil of ultra cold atoms in a Bose-Einstein condensate due to radiation pressure was observed to be compatible with Minkowski momentum.

There is an old measurement of the photon drag effect in semiconductors that agreed with Minkowski's momentum \cite{Gibson1980}, and the reasoning from last paragraph is valid again. This also applies to most cases reported in Ref.~\cite{Strait2019}, where the same effect was measured in thin metal films. However, it is important to notice that one specific measurement in this last reference showed a negative dependence on Minkowski's momentum -- a result that, according to the authors, still lacks theoretical explanation regarding the optical transduction and microscopic momentum exchange mechanisms.

\subsection{Bulk force experiments}

This section deals with the experiments where only the body forces from the striction effects, the first and second terms of Eq.~(\ref{eq:f3}), are relevant. The electrostriction effect was measured inside a fluid in Refs.~\cite{Hakim1962,Zimmerli1999} using high-intensity static fields (the latter one in a microgravity environment) and the results agreed with the so-called Helmholtz force density~\cite{Brevik1979,Helmholtz,Lai1981}, given by $\bb f_{\mathrm{H}}=\varepsilon_0\bm\nabla\left[ \left(\rho_{\mathrm{m}}\partial\varepsilon_{\mathrm{r}}/\partial\rho_{\mathrm{m}}\right)_T |\bb E|^2/2 \right]$. This equation is phenomenological and valid for media in thermodynamic equilibrium with external fields, where $\rho_{\mathrm m}$ denotes mass density and $T$ temperature. We notice that the results from Refs.~\cite{Hakim1962,Zimmerli1999} can also be explained by the MA formulation if we consider the local-field correction in the form of the Clausius-Mossotti relation~\cite{Jackson1999,Aspnes1982}, where we have $\rho_{\mathrm{m}}\partial\varepsilon_{\mathrm{r}}/\partial\rho_{\mathrm{m}}=(\varepsilon_{\mathrm{r}}-1)(\varepsilon_{\mathrm{r}}+2)/3$. 
However, it should be emphasized that the equivalence between MA and Helmholtz force densities is limited to static and quasi-static conditions -- although the mathematical form would be identical, the two force density equations are built under very different assumptions. In particular, it is certainly not expected that the medium is in thermodynamic equilibrium under optical excitation.
Indeed, for optical excitation the local-field correction is typically given as the Lorentz-Lorenz relation~\cite{Kragh2018}. Nonetheless, recent measurements of the electrostriction effect in water for laser excitation at optical frequency were very well described by the MA formulation without the correction to the local-field~\cite{Astrath2022,Astrath2023}. This can be justified by the argumentation presented in Ref.~\cite{Partanen2023}, where this correction is absent due to the optical electrostriction effect causing energetically non-conservative changes in the dipole moments through the variation of the material's local mass density.

A quite intricate measurement of the electromagnetic force inside optical fibers was reported in Ref.~\cite{Xi2021}. It was concluded that the force density has a different symmetry than the expected from MA formulation for this case. It should be mentioned that the irregular position-dependent refractive index in the optical fiber due to its fabrication process may play a significant role in the force density symmetry through the terms proportional to $\bm \nabla \varepsilon$ and $\bm \nabla \mu$. 

In Ref.~\cite{Choi2017} the observation of the Abraham force was reported in a liquid-filled hollow optical fiber, where the Abraham-Minkowski pressure at the free liquid surface was claimed to be carefully suppressed by the geometry of the waveguide. However, striction forces were not considered in the analysis of the results, where it is expected they would generate deformations with contrary direction to the observed one. Furthermore, there would be an additional Abraham-Minkowski force at the ring core/liquid interface, and adhesion effects are also expected to be important -- indeed, these last two forces should partially cancel striction forces. If this cancellation is significant, the remaining force term according to MA formulation would be Abraham's one, in agreement with the authors' conclusions. Alternatively, all the effects but Abraham's force could also be relatively very small. The presence of more than one mode in the excitation wave can also be relevant. 


\subsection{Total force experiments}

As electro- and magnetostriction are gradient forces, they always produce zero total force when integrated over the material's volume, and so they do not contribute to the macroscopic movement of the body~\cite{Brevik1979,Kemp2017}. Thus, the total force acting on a dielectric body should be composed of only the time-derivative term in Eq.~(\ref{eq:f3}), i.e., Abraham's force. The movement of a torsion pendulum induced by the simultaneous application of low frequency time-dependent electric fields and static magnetic fields was measured in Ref.~\cite{Walker1975}, and agreed with Abraham's force. The experiment and results reported in Ref.~\cite{Rikken2012} are similar, but more detailed as it also covered the case of electromagnetic forces generated by static electric fields together with time-varying magnetic fields. At last, in Ref.~\cite{Rikken2011}, the pressure variation of a confined gas due to the presence of electromagnetic fields was measured to be compatible with Abraham's force. All these results are in agreement with MA formulation.

\section{Discussion}\label{sec:disc}

Although, as mentioned earlier, a detailed discussion on the different electromagnetic formulations existing in the literature is out of our scope, we recall that every experiment of the long list cited in the last section can be quantitatively interpreted by using at least one of them -- however, we emphasize that only MA, Einstein-Laub (EL) and Chu formulations can be consistent (at least partially) with all of the experiments. Nevertheless, there are strong theoretical issues with EL and Chu formulations which will be discussed now. As their force density is the same (see Appendix~\ref{sec:app_tab}), we can focus only on the former.

The EL force density can be written as~\cite{Brevik1979}
\begin{eqnarray}\label{eq:f_EL}
    \bb f_{\mathrm{EL}} = \frac{1}{2}\bm\nabla\left(\bb P \!\cdot\! \bb E \right)
 \!+\!\frac{1}{2}\bm\nabla\left(\mu_0\bb M \!\cdot\! \bb H \right)
  \!-\!\frac{1}{2}|\bb E|^2\bm\nabla\varepsilon   \nonumber \\
  \!-\!\frac{1}{2}|\bb H|^2\bm\nabla\mu
  \!+\!\frac{n^2\!-\!1}{c^2}\frac{\partial}{\partial t}\left(\bb E \!\times\! \bb H\right). 
\end{eqnarray}
Comparing the above equation to our proposed force density, Eq.~(\ref{eq:f3}), we see their difference lies in the second term, the magnetostriction effect -- which is a pure magnetic effect.  Such difference is expected because the magnetization mechanism in the EL formulation is due to magnetic monopoles~\cite{EL}, while in MA formulation we adopt the classical Ampèrian current loops model. The magnetic force under the monopole model is~\cite{Boyer1988}
$\bb F_{\mathrm{m}}=(\bb m\cdot\bm\nabla)\bb B$, while for the current loop model we have $\bb F_{\mathrm{l}}=\bm\nabla(\bb m\cdot\bb B)$. Expanding last equation, we have $\bb F_{\mathrm{l}}=(\bb m\cdot\bm\nabla)\bb B+\bb m\times(\bm\nabla\times\bb B)+(\bb B\cdot\bm\nabla)\bb m+\bb B\times(\bm\nabla\times\bb m)$.
Under the point dipole approximation we adopted, the spatial derivatives of $\bb m$ are neglected, so that $\bb F_{\mathrm{l}}=(\bb m\cdot\bm\nabla)\bb B+\bb m\times(\bm\nabla\times\bb B)$ (exactly as in Eq.~(\ref{eq:f_l})). Therefore, the models yield the same force \emph{only if} $\bm\nabla\times\bb B=0$. More importantly, the monopole model has been convincingly observed to disagree with experimental results, while the loop model has shown very good agreement~\cite{Hughes1951,Mezei1986,Jackson1977}. Experimentally, there are unfortunately very few experiments probing the magnetostriction effect~\cite{Samata2000,Wang2018,Ekreem2007,Goldman1951,Klokholm1976,Bellesis1993,Bozorth1953}, and, to our knowledge, no experiment at all under optical excitation -- mainly because thermal effects would in this case be dominant. Additionally, in these works the classical treatment of magnetostriction is phenomenological~\cite{Chikazumi1996,Miyazaki2012}, given in terms of magneto-elastic coefficients, which is a significantly different approach from EL and MA descriptions. Thus, currently EL and MA force densities can not be directly discriminated on experimental basis, and do yield the same results for non-magnetic media. Nonetheless, the major concern about the EL formulation is based on the theoretical and experimental results just discussed, which we feel are compelling enough.

Another important feature of the EL formulation is that it satisfies the so-called duality transformations of electromagnetic fields~\cite{Barnett2015,Jackson1977}. Again, this is expected since in this formulation the polarization and magnetization mechanisms are treated in direct analogy. This is not the case with MA formulation, where these two mechanisms are distinct -- therefore, not satisfying these transformations is actually expected for the MA force density. We notice, however, that a dual-symmetric classical electromagnetic theory can be built without invoking magnetic charges~\cite{Bliokh2013}.


The force density presented in Eq.~(\ref{eq:f3}) must, of course, be theoretically compatible with the covariance requirements from special relativity. It was shown that to fulfil this condition there must be a coupled state of field and matter propagating through the material~\cite{MP1,MP6}, which is described by the so-called mass-polariton (MP) quasi-particle. In this theory, the medium contribution -- whose importance had been already noticed in earlier works in connection to the photon mass drag effect~\cite{Loudon2004,Loudon2005,Mansuripur2005,Milonni2010} -- is described as a mass density wave propagating along with the electromagnetic wave due to the atoms being driven forward by the optical force. This theory was shown to be covariant~\cite{MP4,MP6}, to conserve angular momentum (from orbital and spin origins)~\cite{MP3}, to be consistent with both classical field and photon descriptions~\cite{MP1,MP3} and, at last, to be in agreement with the celebrated recognition of Abraham's and Minkowski's momenta as kinetic and canonical momenta, respectively~\cite{Barnett2010,Barnett2010b}.
The MP dynamics and its momentum transfer employ the Abraham force density, which consists of the three last terms of Eq.~(\ref{eq:f3}). In the present work, the force density additionally has the striction effects, which correspond to pure stresses and do not affect the overall momentum transfer inside the material~\cite{Brevik1979,Partanen2023}. Therefore, we expect Eq.~(\ref{eq:f3}) is also compatible with the MP dynamics, which can potentially provide a complete microscopical description of energy and momentum transfer in linear, lossless and non-dispersive dielectrics, but a detailed discussion will be addressed in a future work. As the fundamental small mass transfer associated to the mass-polariton has not yet been observed, the comparison with the experimental investigations from last section are not affected.

Regarding the \emph{total} forces acting on the medium, it is known that different formulations can consistently provide the same results~\cite{Barnett2006,Loudon2006,Mansuripur2013} -- in fact, it has been argued that, with the proper choice of material contribution, every existing formulation leads to the same results and thus choosing from one of them is just a matter of personal convenience~\cite{Pfeifer2007}. The possibility of experimental discrimination between distinct force densities must, therefore, somehow identify their particular spatio-temporal dependence within the material. This is, of course, a challenging experimental work, but it was reported recently in all-optical pump and probe photo-induced lensing experiments~\cite{Astrath2022,Astrath2023} and the results showed an excellent agreement with MA force density. In fact, these two pioneering works are very valuable as they should change the aforementioned theoretical beliefs that force densities are not unique and that only total forces can be observed.

The new radiation pressure equation for p polarization, Eq.~(\ref{eq:P_p}), was derived for non-magnetic media, where $\bb f_{\mathrm{MA}}$ can be written as $\bb f_{\mathrm{MA}}=(\bb P \cdot \bm \nabla)\bb E+\partial_t \bb P\times \bb B$ (Eq.~(\ref{eq:f}) with $\bb M=0$). Although not in the explicit form of Eq.~(\ref{eq:P_p}), the radiation pressure arising from this force density has already been addressed in the literature~\cite{Loudon2006}, providing consistent theoretical results regarding total momentum conservation. Additionally, we can, for example, take the thought-experiment considered in Ref.~\cite{Mansuripur2005b}, where a p-polarized beam propagating in air enters and then exits a prism's wedges at Brewster angles $\theta_{\mathrm{B}}$ and $\theta_{\mathrm{B}}'$, respectively. In this case, $r_{\mathrm{p}}$ is zero at both interfaces, yielding (apart from a common multiplicative constant) the pressures at entrance and exit interfaces (respectively) $\mathcal{P}_\mathrm{in}^{(\mathrm{p})}=t_{\mathrm{p}}^2(\theta_{\mathrm{B}})\cos^2\theta_{\mathrm{B}}'+t_{\mathrm{p}}^2(\theta_\mathrm{B})\sin^2\theta_{\mathrm{B}}'/2+\sin^2\theta_{\mathrm{B}}/2$ and $\mathcal{P}_{\mathrm{out}}^{(\mathrm{p})}=-[t_{\mathrm{p}}^2(\theta_{\mathrm{B}}')\cos^2\theta_{\mathrm{B}}+t_{\mathrm{p}}^2(\theta_{\mathrm{B}}')\sin^2\theta_{\mathrm{B}}/2+\sin^2\theta_{\mathrm{B}}'/2]t_{\mathrm{p}}^2(\theta_{\mathrm{B}})$, where $\theta_{\mathrm{B}} = \tan^{-1}(n_2/n_1)$ and $\theta_{\mathrm{B}}+\theta_{\mathrm{B}}'=90^{\circ}$. It can be numerically verified that the sum of these pressures is zero, as required for conservation of total momentum in this situation -- in fact, this result still holds regardless of the dielectric medium surrounding the prism. 
For general incidence angles, we must resort to computational techniques as the internal beam reflections will naturally make the analysis much more complicated.

Though our presented derivation of the electromagnetic force density is currently limited to the simplest type of dielectric materials, we notice that there are many possibilities that can be explored starting from it. One can, for instance, try to extend the theory to more complex materials, where effects such as dispersion, absorption, anisotropy and nonlinearities can take place. Consideration of non-conservative optical forces can also be of interest~\cite{Sukhov2017}. 
The analysis of angular momentum distributions inside materials is also relevant, especially because light can have both spin and orbital angular momentum~\cite{Bliokh2015} -- this has been simulated, for example, in Ref.~\cite{MP2}. 
Additionally, one can search for a physically more fundamental formalism by fully working in the Quantum Mechanics regime, where effects such as field fluctuations and vacuum pressure can occur -- there are already some theoretical works in this regard, contemplating, for example, the radiation pressure due to the quantum-mechanical Lorentz force~\cite{Loudon2002}, QED corrections to the Abraham force~\cite{LeFournis2022} and Casimir-like effects~\cite{Feigel2004}.

\section{Conclusions}\label{sec:conc}

In this work we presented a new equation for the electromagnetic force densities inside linear, isotropic, non-dispersive and lossless dielectric material, called Microscopic Ampère formulation. This result is derived from the well-established dipolar approximation for electromagnetic sources. Among the other formulations existing in the literature, our proposal is the only one capable of explaining the vast majority of experimental works reported to date and to be simultaneously in accordance with the absence of magnetic monopoles in nature. 
Additionally, the proposed force density needs the inclusion of the hidden momentum contribution, whose origin was briefly discussed and clarified in the context of the Abraham-Minkowski controversy.
A new expression for the radiation pressure in non-magnetic dielectrics under oblique incidences for p-polarized beams was also derived. It is consistent with momentum conservation, but more investigations are necessary -- especially experimental ones.

Even though currently limited to classical Physics and to the simplest type of dielectric materials, the electromagnetic force density and pressure presented here cover a lot of practical applications and provide an important step towards obtaining the definitive knowledge of the behaviour of light inside matter.

\begin{acknowledgments}
The research leading to these results received funding from CNPq (304738/2019-0), CAPES (Finance Code 001), Funda\c{c}\~{a}o Arauc\'{a}ria, and FINEP. M.P. acknowledges funding from the Academy of Finland under Contract No. 349971.
\end{acknowledgments}

\appendix
\onecolumngrid
\section{Force densities according to different formulations}\label{sec:app_tab}
For a modern derivation of the electromagnetic force densities listed below, we refer to the works~\cite{Brevik1979,Milonni2010,LL,Anghinoni2022,Chu}. 

\begin{table*}[!h]
\centering
\begin{threeparttable}
\renewcommand{\arraystretch}{1.6}
\caption{Force density according to different electromagnetic formulations.}
\label{tab:set_tab}
\begin{tabular}{ll}
\hline\hline
\textbf{Formulation}  & \quad\quad\quad\quad\quad\quad\quad\quad\textbf{Force density} \\ 
\hline
Minkowski & $\bb f_{\mathrm{M}} = -\frac{1}{2}|\bb E|^2\bm\nabla\varepsilon-\frac{1}{2}|\bb H|^2\bm\nabla\mu$    \\
\hline
Abraham    & $\bb f_{\mathrm{Ab}}=  -\frac{1}{2}|\bb E|^2\bm\nabla\varepsilon-\frac{1}{2}|\bb H|^2\bm\nabla\mu+\frac{n^2-1}{c^2}\frac{\partial(\bb E \times \bb H)}{\partial t}$ \\ 
\hline
Conventional Ampère 
& $\bb f_{\mathrm{A}} \!=\! -(\bm\nabla\!\cdot\! \bb P)\bb E+(\frac{\partial \bb P}{\partial t}\!+\!\bm\nabla\!\times\!\bb M)\!\times\! \bb B $ \\  
\hline
Einstein-Laub  & $\bb f_{\mathrm{EL}} = \frac{1}{2}\bm\nabla(\bb P \cdot\bb E)+\frac{1}{2}\bm\nabla(\mu_0\bb M\cdot \bb H) -\frac{1}{2}|\bb E|^2\bm\nabla\varepsilon-\frac{1}{2}|\bb H|^2\bm\nabla\mu+\frac{n^2-1}{c^2}\frac{\partial(\bb E \times \bb H)}{\partial t}$ \\
\hline
Chu\tnote{a} & $\bb f_{\mathrm{C}} = \frac{1}{2}\bm\nabla(\bb P \cdot\bb E)+\frac{1}{2}\bm\nabla(\mu_0\bb M\cdot \bb H) -\frac{1}{2}|\bb E|^2\bm\nabla\varepsilon-\frac{1}{2}|\bb H|^2\bm\nabla\mu+\frac{n^2-1}{c^2}\frac{\partial(\bb E \times \bb H)}{\partial t}$\\
\hline
Helmholtz\tnote{b} &$\bb f_{\mathrm{H}}=-\frac{1}{2}|\bb E|^2\bm\nabla\varepsilon-\frac{1}{2}|\bb H|^2\bm\nabla\mu+\frac{1}{2}\bm\nabla\left[|\bb E|^2\rho_\mathrm{m}\left(\frac{\partial \varepsilon}{\partial \rho_\mathrm{m}}\right)_T\right]+\frac{1}{2}\bm\nabla\left[|\bb H|^2\rho_\mathrm{m}\left(\frac{\partial \mu}{\partial \rho_\mathrm{m}}\right)_T\right]$\\
\hline\hline
\end{tabular}
\begin{tablenotes}\footnotesize
\item [a] Considering the total stress-energy tensor of field plus matter.
\item [b] Valid for static or quasi-static fields.
\end{tablenotes}
\end{threeparttable}
\end{table*}
\twocolumngrid

\section{Lagrangian dynamics for homogeneous media}\label{sec:Lag}
The force density from Eq.~(\ref{eq:f3}) can be derived for homogeneous media (where $\bm\nabla\varepsilon=\bm\nabla\mu=0$) from the Lagrangian approach as well. The potential energy of induced electric and magnetic dipoles is known to be~\cite{Jackson1999}
\begin{equation}
    U = -\frac{1}{2}\bb p\cdot \bb E-\frac{1}{2}\bb m\cdot \bb B.
\end{equation}

In the presence of an external electromagnetic force, the dipoles are put into motion, acquiring a velocity $\bb v$ given in the laboratory frame. Due to this motion, to first order in $|\bb v|/c$ the dipole moments are given in the laboratory as $\bb p'=\bb p+\bb v\times\bb m/c^2$ and $\bb m'=\bb m -\bb v \times \bb p$~\cite{Hnizdo2012}. Therefore, in the laboratory frame the non-relativistic Lagrangian reads
\begin{eqnarray}
    L &=& m v^2 + \frac{1}{2}\bb p'\cdot\bb E+\frac{1}{2}\bb m' \cdot \bb B \nonumber\\
    &=& m v^2\!+\!\frac{1}{2}\left(\bb p\!+\!\frac{1}{c^2}\bb v \!\times\! \bb m\right)\!\cdot\bb E\!\nonumber\\
    &+&\!\frac{1}{2}\left(\bb m\!-\!\bb v \!\times\! \bb p\right)\!\cdot\bb B,
\end{eqnarray}
where the fields $\bb E$ and $\bb B$ are assumed to be also given in the laboratory frame. The canonical momentum is then
\begin{eqnarray}
    \frac{\partial L}{\partial \bb v}=m \bb v -\bb p\times\bb B + \frac{1}{c^2}\bb m \times \bb E, 
\end{eqnarray}
from where we identify the Röntgen and hidden momentum contributions (the two last terms, respectively), discussed in Sec.~\ref{sec:HM}. The force $m \dot{\bb v}$ on the dipole is then obtained from the Euler-Lagrange equation, 
\begin{eqnarray}
    \frac{\dd}{\dd t}\frac{\partial L}{\partial \bb v} = \frac{\partial L}{\partial \bb r},
\end{eqnarray}
which yields
\begin{eqnarray}
  m \dot{\bb v} +\frac{\dd}{\dd t}\left(-\bb p\times\bb B + \frac{1}{c^2}\bb m \times \bb E\right) = \nonumber\\
  \frac{1}{2}\bm \nabla \left [\left(\bb p+\frac{1}{c^2}\bb v \times \bb m\right)\cdot\bb E+\left(\bb m-\bb v \times \bb p\right)\cdot\bb B\right]. 
\end{eqnarray}
Isolating $m \dot{\bb v}$ and neglecting the terms linear in $\bb v$, we obtain
\begin{eqnarray}
    m \dot{\bb v} =  \frac{1}{2}\bm \nabla \!\left(\bb p \!\cdot\! \bb E \!+\! \bb m\!\cdot\! \bb B\right)\!+\!\frac{\partial}{\partial t}\left(\bb p\!\times\!\bb B\!-\!\frac{1}{c^2}\bb m\!\times\!\bb E\right)\!,
\end{eqnarray}
which, once divided by the volume and rearranged, leads to Eq.~(\ref{eq:f3}) for homogeneous media. Notice that a similar Lagrangian approach was reported in Ref.~\cite{Hnizdo2012b} for static fields and permanent dipoles -- therefore, the derivation shown here is an extension to induced dipoles and time-dependent fields. This approach is here convenient as it elucidates very clearly the origin of the Röntgen and hidden momentum contributions as transformations between inertial reference frames of the moving dipole moments -- however, it does not address the possibility of a non-homogeneous medium, which is crucial to explain the radiation pressure experiments, as seen in Sec.~\ref{sec:exp_p}.

\section{Force density calculation}\label{sec:app_f}

The following derivation can also be found in the Supplementary Material of Ref.~\cite{Astrath2022}. Here, we add it with some more details for completeness and convenience.

We start from the Lorentz force equation with the microscopic sources, namely
\begin{eqnarray}\label{eq:F}
\bb F = \int_{\delta V} \left(\left[-(\bb p\cdot\bm\nabla)\delta^3(\bb r)\right]\bb E + \dot{\bb p}\delta^3(\bb r)\times \bb B-\right.\nonumber\\
\left.\left[(\bb m \!\times\! \bm \nabla )\delta^3(\bb r)\right]\!\times\! \bb B\right) \!\dd^3\bb r.    
\end{eqnarray}
As stated in Sec.~\ref{sec:f+p}, the integral is performed over a small volume of the dielectric, $\delta V$, which is microscopically large but still much smaller than the dielectric's macroscopic volume.
We must now calculate Eq.~(\ref{eq:F}), which gives the total force on the dielectric. Each term of the integral will be treated separately. The first term is
\begin{equation}
    \bb F_1 = -\int [(\bb p\cdot\bm\nabla) \delta^3(\bb r)]\bb E \dd^3\bb r.
\end{equation}

Writing explicitly the components of the term inside the brackets, we have
\begin{eqnarray}
   \bb F_1 = -\int [p_x\delta(y)\delta(z)\delta'(x)+p_y\delta(x)\delta(z)\delta'(y)\nonumber\\
   + p_z\delta(x)\delta(y)\delta'(z)]\bb E \dd^3\bb r.
\end{eqnarray}

Integrating each term by parts we obtain
\begin{eqnarray}
    \bb F_1 &=& \partial_x(p_x\bb E)+\partial_y(p_y\bb E)+\partial_z(p_z\bb E) \nonumber \\
    &=& (\bb p\cdot\bm\nabla)\bb E + (\bm\nabla\cdot\bb p)\bb E.
\end{eqnarray}
The last term of the above equation is related to the spatial extent of the dipole and under the point dipole approximation can be neglected as $\bb p$ does not depend on $\bb r$ -- so, $\bb F_1 = (\bb p\cdot\bm\nabla)\bb E$.

The second term of Eq.~(\ref{eq:F}) is integrated trivially:
\begin{equation}
    \bb F_2 = \int [\bb{\dot p}\delta^3(\bb r)]\times\bb B\dd ^3\bb r = \bb{\dot p} \times \bb B.
\end{equation}

The last term of Eq.~(\ref{eq:F}) is
\begin{eqnarray}
    \bb F_3 &=& -\int [(\bb m \times \bm \nabla )\delta^3(\bb r)]\times \bb B \dd^3\bb r \nonumber \\
    &=& -\int \{[\delta(x)\delta(y)m_y\delta'(z)-\delta(x)\delta(z)m_z\delta'(y)]\bb {\hat i}\nonumber\\
    &+&[\delta(y)\delta(z)m_z\delta'(x)-\delta(x)\delta(y)m_x\delta'(z)]\bb{\hat j}\nonumber \\
   &+& [\delta(x)\delta(z)m_x\delta'(y)-\delta(y)\delta(z)m_y\delta'(x)]\bb{\hat k}\}\times \bb B \dd^3\bb r \nonumber \\
   &=&-\int \bb C \times \bb B \dd^3\bb r,
\end{eqnarray}
where the auxiliary vector $\bb C$ has been implicitly defined for simplicity and $\bb{\hat i},\bb{\hat j},\bb{\hat k}$ denote the cartesian unit vectors. Performing the cross product, we have
\begin{eqnarray}
    \bb F_3 = -\int[(C_y B_z-C_z B_y)\bb{\hat i}+(C_z B_x - C_x B_z)\bb{\hat j}\nonumber\\
    +(C_x B_y - C_y B_x)\bb{\hat k}] \dd^3\bb r.
\end{eqnarray}
The $x$ component of $\bb F_3$ is then
\begin{eqnarray}
    F_{3,x} &=& -\int [\delta(y)\delta(z)m_z\delta'(x)B_z-\delta(x)\delta(y)m_x\delta'(z)B_z\nonumber \\
    &-& \delta(x)\delta(z)m_x\delta'(y)B_y+\delta(y)\delta(z)m_y\delta'(x)B_y] \dd^3\bb r \nonumber \\
    &=& m_z\partial_x B_z\!-\!m_x\partial_z B_z \!-\! m_x\partial_y B_y\!+\!m_y\partial_x B_y,
\end{eqnarray}
where we have integrated by parts again and applied the point dipole approximation. The $y$ and $z$ components are obtained analogously, yielding
\begin{eqnarray}
    F_{3,y} \!=\! m_x\partial_y B_x\!-\!m_y\partial_x B_x \!-\! m_y\partial_z B_z+m_z\partial_y B_z, \\
    F_{3,z} \!=\! m_y\partial_z B_y\!-\!m_z\partial_y B_y \!-\! m_z\partial_x B_x+m_x\partial_z B_x.
\end{eqnarray}

To obtain a more compact result, notice that
\begin{eqnarray}
    F_{3,x} - [\bb m \!\times\! (\bm \nabla \!\times\! \bb B)]_x - [(\bb m\!\cdot\!\bm\nabla)\bb B]_x = \nonumber \\
    -m_x \partial_z B_z - m_x \partial_y B_y + m_y \partial_y B_x
    +  m_z \partial_z B_x\nonumber \\ - m_x \partial_x B_x -m_y \partial_y B_x -m_z \partial_z B_x \nonumber \\
    =-m_x(\bm\nabla\cdot\bb B) = 0.
\end{eqnarray}
This means that $ F_{3,x} = [\bb m \times (\bm \nabla \times \bb B)]_x + [(\bb m\cdot\bm\nabla)\bb B]_x$, and, consequently, 
\begin{equation}\label{eq:f_l}
   \bb F_3 = \bb m \times (\bm \nabla \times \bb B) + (\bb m\cdot\bm\nabla)\bb B, 
\end{equation}
which completes our calculation.

Summing the three contributions to the force and dividing by the dielectric's volume, we obtain the position- and time-dependent force density in the so called Microscopic Ampère (MA) formulation
\begin{eqnarray}
\label{eq:f}
    \bb f_{\mathrm{MA}}=\left(\bb P\cdot \bm \nabla \right) \bb E+ \dot{\bb{P}} \times \bb B + \bb M \times \left(\bm \nabla \times \bb B\right)\nonumber \\
    +\left(\bb M  \cdot\bm\nabla \right) \bb B,
\end{eqnarray}
where the fields are the macroscopic ones evaluated at the location of the dipoles. 

Eq.~(\ref{eq:f}) was first given, to our knowledge, in Ref.~\cite{Griffiths2015} -- but it was not explored in the context of the Abraham-Minkowski controversy. It is compatible with a pure, point dielectric dipole located at the origin and at rest in its own frame. 
Explicitly, the time derivative term is
\begin{equation}\label{eq:f1}
    \dot{\bb{P}} \times \bb B = \frac{\partial \bb P}{\partial t} \times \bb B,
\end{equation}
as the dipole is assumed to be at rest.

Now, we want to rewrite Eq.~(\ref{eq:f}) in a form which it is most conveniently interpreted in the context of the Abraham-Minkowski controversy. This task will require a lot of vector algebra. We start by using the vector property $\bm \nabla \left(\bb U  \cdot \bb V\right) =  (\bb U \cdot\bm \nabla)\bb V +  (\bb V \cdot\bm \nabla)\bb U+\bb U \times\left(\bm \nabla \times \bb V\right)+\bb V \times \left(\bm \nabla \times \bb U\right)$, which allows us to write
\begin{eqnarray}\label{eq:pol}
    \left(\bb P\cdot \bm \nabla \right) \bb E = \bm \nabla\left(\bb P \cdot \bb E\right)-\left(\bb E\cdot \bm \nabla \right) \bb P\nonumber \\
    -\bb E \times \left(\bm \nabla \times \bb P\right)-\bb P \times \left(\bm \nabla \times \bb E\right)
\end{eqnarray}
and
\begin{eqnarray}\label{eq:mag}
  \bb M \times \left(\bm \nabla \times \bb B\right)+\left(\bb M  \cdot\bm\nabla \right) \bb B =  \bm \nabla\left(\bb M\cdot \bb B\right) \nonumber \\
  - \bb B \times \left(\bm \nabla \times \bb M\right)-\left(\bb B  \cdot\bm\nabla \right) \bb M.  
\end{eqnarray}

For linear isotropic media, the medium responses are given by $\bb P = \varepsilon_0\chi_{\mathrm{e}}\bb E$ and $\bb M = \chi_{\mathrm{m}}\bb H$, where $\chi_{\mathrm{e}}$ and $\chi_{\mathrm{m}}$ are the electric and magnetic susceptibilities, respectively. Working on Eq.~(\ref{eq:pol}), we have
\begin{eqnarray}\label{eq:pol1}
     \left(\bb P\cdot \bm \nabla \right) \bb E = \varepsilon_0\bm\nabla\left(\chi_{\mathrm{e}}|\bb E|^2\right)
     -\varepsilon_0\left(\bb E\cdot\bm\nabla\right)\left(\chi_{\mathrm{e}}\bb E\right)\nonumber\\ 
     -\varepsilon_0\bb E \times\left(\bm \nabla \times (\chi_{\mathrm{e}} \bb E)\right)    -\varepsilon_0\chi_{\mathrm{e}}\bb E\times\left(\bm \nabla \times \bb E\right).
\end{eqnarray}
The first term on the right hand side of this equation is
\begin{equation}
    \varepsilon_0\bm\nabla\left(\chi_{\mathrm{e}}|\bb E|^2\right) = \varepsilon_0|\bb E|^2\bm \nabla\chi_{\mathrm{e}} + \varepsilon_0\chi_{\mathrm{e}}\bm\nabla |\bb E|^2. 
\end{equation}
The second term is
\begin{eqnarray}
  \varepsilon_0\left(\bb E\cdot\bm\nabla\right)\left(\chi_{\mathrm{e}}\bb E\right) = \varepsilon_0\chi_{\mathrm{e}}\left(\bb E \cdot \bm \nabla\right)\bb E+\nonumber\\
  \varepsilon_0\left(\bb E\cdot \bm \nabla \chi_{\mathrm{e}}\right)\bb E.  
\end{eqnarray}
The third term is
\begin{eqnarray}
 \varepsilon_0\bb E \!\times\!\left(\bm \nabla \!\times\! (\chi_{\mathrm{e}} \bb E)\right) &=& \varepsilon_0 \bb E \!\times\! \left(\chi_{\mathrm{e}}\bm\nabla\!\times\!\bb E + (\bm\nabla\chi_{\mathrm{e}})\!\times\!\bb E\right) \nonumber    \\ 
 &=& -\bb P \times \frac{\partial \bb B}{\partial t}+\varepsilon_0|\bb E|^2\bm\nabla\chi_{\mathrm{e}}\nonumber\\
&-&\varepsilon_0\left(\bb E  \cdot\bm\nabla\chi_{\mathrm{e}}\right)\bb E,
\end{eqnarray}
where the vector property $\bb a \times (\bb b \times \bb c)=(\bb a \cdot \bb c)\bb b-(\bb a \cdot \bb b)\bb c$ was used.

The last term on Eq.~(\ref{eq:pol1}) can be described by Faraday's law as well, so that
\begin{eqnarray}
  \left(\bb P\cdot \bm \nabla \right) \bb E \!&=& \!
  \varepsilon_0|\bb E|^2\bm \nabla\chi_{\mathrm{e}} \!+\! \varepsilon_0\chi_{\mathrm{e}}\bm\nabla |\bb E|^2
  \!-\!\varepsilon_0\chi_{\mathrm{e}}\left(\bb E \cdot \bm \nabla\right)\bb E  \nonumber\\
   &-&\varepsilon_0\left(\bb E\cdot \bm \nabla \chi_{\mathrm{e}}\right)\bb E
  +\bb P \times \frac{\partial \bb B}{\partial t}
  -\varepsilon_0|\bb E|^2\bm\nabla\chi_{\mathrm{e}} \nonumber \\  &+&\varepsilon_0\left(\bb E  \cdot\bm\nabla\chi_{\mathrm{e}}\right)\bb E
  +\bb P \times \frac{\partial \bb B}{\partial t}.
\end{eqnarray}

Simplifying the last equation, we obtain
\begin{equation}
\left(\bb P\cdot \bm \nabla \right) \bb E = 
 \frac{\varepsilon_0(\varepsilon_{\mathrm{r}}-1)}{2}\bm\nabla |\bb E|^2+\bb P \times \frac{\partial \bb B}{\partial t},
\end{equation}
where we used $\bm \nabla |\bb E|^2/2=(\bb E\cdot\bm\nabla)\bb E+\bb E\times(\bm\nabla\times\bb E)$.

Adding Eq.~(\ref{eq:f1}) to last equation yields
\begin{equation}\label{eq:pol2}
\left(\bb P\cdot \bm \nabla \right) \bb E + \bb{\dot P}\!\times\!\bb B = 
 \frac{\varepsilon_0(\varepsilon_{\mathrm{r}}-1)}{2}\bm\nabla |\bb E|^2
 +\frac{\partial}{\partial t}\left(\bb P \!\times\! \bb B\right).
\end{equation}

Proceeding analogously for Eq.~(\ref{eq:mag}), we have
\begin{eqnarray}
  \bb M \times \left(\bm \nabla \!\times\! \bb B\right)+\left(\bb M  \cdot\bm\nabla \right) \bb B =
  \bm \nabla \left(\bb \chi_{\mathrm{m}} \bb H\cdot \mu \bb H\right) \nonumber \\
  - \mu\bb H \times \left(\bm \nabla \times \bb \chi_{\mathrm{m}}\bb H\right)-\left(\mu\bb H  \cdot\bm\nabla \right) \left(\chi_{\mathrm{m}}\bb H\right),
\end{eqnarray}
where $\bb B = \mu_0 (\bb H+\bb M) = \mu \bb H$ was used.

The first term on the right hand side of last equation is
\begin{equation}
 \bm \nabla \left(\chi_{\mathrm{m}} \bb H\cdot \mu \bb H\right) =   (2\mu_{\mathrm{r}}\!-\!1)|\bb H|^2\bm\nabla\mu +\mu(\mu_{\mathrm{r}}\!-\!1)\bm\nabla|\bb H|^2.   
\end{equation}

The second term is
\begin{eqnarray}
\mu\bb H \!\times\! \left(\bm \nabla \!\times\! \bb \chi_{\mathrm{m}}\bb H\right)  \!&=&\!   \mu \bb H \times \left(\chi_{\mathrm{m}}\bm \nabla\times\bb H+(\bm\nabla\chi_{\mathrm{m}})\times\bb H\right) \nonumber \\ 
&=&\mu\chi_{\mathrm{m}}\bb H\!\times\!\left(\bm\nabla\!\times\!\bb H\right)+\mu \bb H\!\times\!\left(\bm\nabla\chi_{\mathrm{m}}\! \times\! \bb H\right) \nonumber \\ 
&=& \mu\chi_{\mathrm{m}}\bb H\!\times\!\left(\bm\nabla\!\times\!\bb H\right)+\mu |\bb H|^2\bm\nabla\chi_{\mathrm{m}}\nonumber \\
\!&-&\!\left(\mu\bb H\!\cdot\!\bm\nabla\chi_{\mathrm{m}}\right)\bb H.
\end{eqnarray}

The third term is
\begin{equation}
 \left(\mu\bb H  \!\cdot\!\bm\nabla \right) \left(\chi_{\mathrm{m}}\bb H\right) =  \mu \left(\bb H\!\cdot\!\bm\nabla\chi_{\mathrm{m}}\right)\bb H+\mu\chi_{\mathrm{m}}\left(\bb H\!\cdot\!\bm\nabla\right)\bb H. \end{equation}

Summing the three terms, Eq.~(\ref{eq:mag}) becomes
\begin{eqnarray}
  \bb M \!\times\! \left(\bm \nabla \!\times\! \bb B\right)+\left(\bb M  \!\cdot\!\bm\nabla \right) \bb B =\nonumber\\ \mu(\mu_{\mathrm{r}}\!-\!1)\bm\nabla|\bb H|^2  - \mu(\mu_{\mathrm{r}}\!-\!1)\left(\bb H\!\cdot\!\bm\nabla\right)\!\bb H \nonumber\\
  -\mu(\mu_{\mathrm{r}}\!-\!1)\bb H\!\times\!\left(\bm\nabla\!\times\!\bb H\right)\!
  +(\mu_{\mathrm{r}}\!-\!1)|\bb H|^2\bm\nabla\!\mu 
 ,
\end{eqnarray}
which, using $\bm \nabla |\bb H|^2/2=(\bb H\cdot\bm\nabla)\bb H+\bb H\times(\bm\nabla\times\bb H)$,  is simplified to
\begin{eqnarray}\label{eq:mag2}
  \bb M \times \left(\bm \nabla \! \times\! \bb B\right)+\left(\bb M  \!\cdot\!\bm\nabla \right) \bb B = (\mu_{\mathrm{r}}\!-\!1)|\bb H|^2\bm\nabla\mu \nonumber\\
  +\frac{\mu(\mu_{\mathrm{r}}-1)}{2}\bm\nabla|\bb H|^2.  
\end{eqnarray}

The force density in linear media is then the sum of Eqs.~(\ref{eq:pol2}) and (\ref{eq:mag2}), namely
\begin{eqnarray}\label{eq:f2}
    \bb f_{\mathrm{MA}} = \frac{\varepsilon_0(\varepsilon_{\mathrm{r}}-1)}{2}\bm\nabla |\bb E|^2
     +(\mu_{\mathrm{r}}-\!1)|\bb H|^2\bm\nabla\mu \nonumber\\
     +\!\frac{\mu(\mu_{\mathrm{r}}-1)}{2}\bm\nabla|\bb H|^2 
 +\frac{\partial}{\partial t}\left(\bb P \!\times \!\bb B\right). 
\end{eqnarray}

\section{Radiation pressure at oblique incidence}\label{sec:app_P}

The radiation pressure for oblique incidence in non-magnetic dielectrics is given in literature by Eq.~(\ref{eq:P_lit}). This equation is equivalent to~\cite{Hallanger2005}
\begin{eqnarray}\label{eq:P_app}
    \mathcal{P}_{\mathrm{rad}} &=& - \frac{I}{2 c}\frac{n_2^2-n_1^2}{n_2}\frac{\cos \theta_{\mathrm{i}}}{\cos \theta_{\mathrm{t}}}\left[(\sin^2 \theta_{\mathrm{i}}
    + \cos^2 \theta_{\mathrm{t}})T_{\mathrm{p}}\cos^2\alpha \right.\nonumber\\  
    &&\left.+ T_{\mathrm{s}}\sin^2\alpha\right],
\end{eqnarray}
where $\alpha$ is the angle between the electric field and the plane of incidence. Thus, for $\alpha = 0$ we have a p polarized beam, while for $\alpha = \pi/2$ we have a s polarized beam. In the former case, we have then
\begin{equation}
    \mathcal{P}_{\mathrm{rad}}^{(\mathrm{p})} = - \frac{I}{2 c}\frac{n_2^2-n_1^2}{n_2}\frac{\cos \theta_{\mathrm{i}}}{\cos \theta_{\mathrm{t}}}\left[(\sin^2 \theta_{\mathrm{i}} + \cos^2 \theta_{\mathrm{t}})T_{\mathrm{p}}\right].
\end{equation}

We can rewrite this equation by using the relation $T = (n_2\cos\theta_{\mathrm{t}}/n_1\cos\theta_{\mathrm{i}})|t^2|$, which is valid for both polarizations~\cite{Hecht}, obtaining
\begin{equation}
    \mathcal{P}_{\mathrm{rad}}^{(\mathrm{p})} = - \frac{n_1 I}{2 c}\frac{n_2^2-n_1^2}{n_1}\left[(\sin^2 \theta_{\mathrm{i}} + \cos^2 \theta_{\mathrm{t}})t^2_{\mathrm{p}}\right].
\end{equation}
Recalling that the (instantaneous) intensity for a plane wave is $I = \varepsilon_0 c n E_0^2$ we have then
\begin{equation}\label{eq:Pp_lit}
    \mathcal{P}_{\mathrm{rad}}^{(\mathrm{p})} = - \frac{(\varepsilon_2-\varepsilon_1)}{2}E_0^2\left[(\sin^2 \theta_{\mathrm{i}} + \cos^2 \theta_{\mathrm{t}})t^2_{\mathrm{p}}\right].
\end{equation}

By applying Snell's law and the relation $n_1^2 E_{z,\mathrm{i}} = n_2^2 E_{z,\mathrm{t}}$ to the last equation, it is possible to write it explicitly in terms of the field components as
\begin{equation}    \mathcal{P}_{\mathrm{rad}}^{(\mathrm{p})} = - \frac{(\varepsilon_2-\varepsilon_1)}{2}\left(E_x^2+E_y^2+E_{z,\mathrm{i}}E_{z,\mathrm{t}}\right).
\end{equation}
The use of the last term in the above equation as the average of the squared field component normal to the interface is not physically justified. Therefore, we expect that Eq.~(\ref{eq:P_p}) is the correct one for radiation pressure due to p-polarized beams. For s-polarized beams, Eqs.~(\ref{eq:P_app}) and (\ref{eq:P_lit}) yield the same results.

\bibliography{Refs} 

\end{document}